\documentclass[]{aa}
\usepackage{graphicx,txfonts,color,natbib}
\bibpunct{(}{)}{;}{a}{}{,}
\setcitestyle{comma}

\begin{document}
\title{Interaction of the massive cluster system Abell 3016/3017 embedded in a cosmic filament}
\author{Gayoung Chon\inst{1,2}, Hans B\"ohringer\inst{1,2}, Sarthak Dasadia\inst{3}, Matthias Kluge\inst{2}, Ming Sun\inst{3}, William R. Forman\inst{4}, Christine Jones\inst{4}}
\institute{
$^1$ Max-Planck-Institut f\"ur extraterrestrische Physik,
D-85748 Garching, Germany\\
$^2$ Universit\"ats-Sternwarte M\"unchen, Fakult\"at f\"ur Physik,
Ludwig-Maximilian-Universit\"at M\"unchen, Scheinerstr. 1, D-81679 M\"unchen, Germany \\
$^3$ Department of Physics \& Astronomy, University of Alabama in Huntsville, Huntsville, AL 35899, USA \\
$^4$ Harvard-Smithsonian Center for Astrophysics, 60 Garden Street, Cambridge, MA 02138, USA
}

\date{Received 21 March 2018 / Accepted 5 November 2018}
\abstract{
The galaxy cluster system RXCJ0225.9-4154 with the two sub-clusters A3016 and A3017 is embedded
in a large-scale structure filament with signatures of filamentary accretion.
In a Chandra observation of this system at a redshift of $z$=0.2195 we detect both clusters in X-rays.
In addition we detect a filament of X-ray emission connecting the two clusters and a galaxy group therein.
The main cluster, A3017, shows indications of shocks most probably from a recent
 interaction with cluster
components along the filament axis as well as a cold front at about 150~kpc from
 the cluster centre.
The filament between the two clusters is likely to be heated by the accretion sh
ocks of the clusters.
We discuss two scenarios for the origin of the X-ray filament between the two cl
usters.
In the first scenario the material of the filament has been ripped off of A3017
during
the fly-by of A3016 and is now trailing the latter sub-cluster.
Support for this scenario is a gas deficit on the eastern side of A3017.
In the second scenario the filament between the two clusters does not come from
either of them, but
a significant contribution could come from the galaxy group located inside and t
he entire structure
is on its first collapse.
We favour the second explanation as the gas mass in the filament seems to be too
 large to be supplied
by the interaction of the two Abell clusters.
The paper describes many properties of the components of this cluster merger sys
tem that
are used to assist the interpretation of the observed configuration.
}

\authorrunning{Chon et al.}
\titlerunning{Merging cluster system embedded in a cosmic filament}
\maketitle
%
%________________________________________________________________
%
\section{Introduction}
The currently adopted model for the formation of the large-scale structure of the universe,
as also studied in detail with cosmological N-body simulations, displays the emergence
of a cosmic web of filamentary structures, where galaxy clusters grow preferentially
at the intersections of the filaments.
The dominant mode of growth of clusters is the inhomogeneous accretion of matter from these filaments. 
Because the way galaxy clusters accrete material from their surroundings is a determining factor 
for the resulting structure of the clusters, in particular for the thermodynamics of their 
intra-cluster medium (ICM), it is important to also study this process observationally. 
In X-ray observations, which provide much detail about the structure of clusters of galaxies, 
we have only seen this clearly in a few cases of this connection of clusters to the cosmic web 
(e.g.~\citet{kull99,scharf00,werner08,buote09,eckert15}).

We found an interesting cluster system, RXCJ0225.9-4154, which is embedded in a large-scale 
galaxy filament and shows an interesting dynamical behaviour in X-rays. 
RXCJ0225.9-4154 (A3016/3017) was identified as an X-ray luminous cluster system in our REFLEX 
cluster survey based on the ROSAT All-Sky Survey (RASS). 
At redshift $z$=0.2195 with an X-ray luminosity of $L_X=6.5\times10^{44}$~erg~s$^{-1}$ (0.1--2.4 keV) 
it is one of the eleven most luminous clusters at $z \le$~0.22 in the southern sky~\citep{boehringer13}.
It also belongs to the REFLEX superstes-cluster sample~\citep{chon13} which implies that it is located 
in an overdense region compared to the mean density at its redshift. 

In this paper we use two archival exposures from Chandra ACIS-I observations to study the properties 
of the cluster system and to infer the formation history.
Our analysis of the Chandra data shows a complex structure of the cluster system with two sub-clusters, 
A3017 and A3016, connected by an X-ray luminous bridge aligned along the larger galaxy filament with 
signatures of cluster and filament interactions. 
We report in the paper on a detailed study of the X-ray observation of the system and offer our 
interpretations, which we also compare to the results produced by~\citet{parekh}.

In Sect. 2 we describe the X-ray data and give a brief summary of the analysis and Sect. 3 provides 
an overview of the morphology of the system in X-rays, while Sect. 4 reports the properties and 
masses of the sub-cluster components. 
In Sect. 5 we analyse the properties of the filamentary structures connecting the two sub-clusters 
and the clusters with the filament, and in Sect. 6 we conclude on the overall dynamics of the system 
and compare our findings and interpretations to those of~\citet{parekh}. 
Sect. 7 provides a summary. 

For the derivation of distance-dependent parameters we use a geometrically flat $\Lambda$CDM model 
with $\Omega_M = 0.3$ and $h_{70} = H_0/70$ km s$^{-1}$~Mpc$^{-1}$ = 1.
All uncertainties without further specifications refer to 1$\sigma$ confidence limits. 
At the cluster distance, 1 arcsec corresponds to the physical scale of 3.55~kpc for the adopted cosmology.

\section{Observations and data reduction}

\begin{figure}
  \includegraphics[width=\columnwidth]{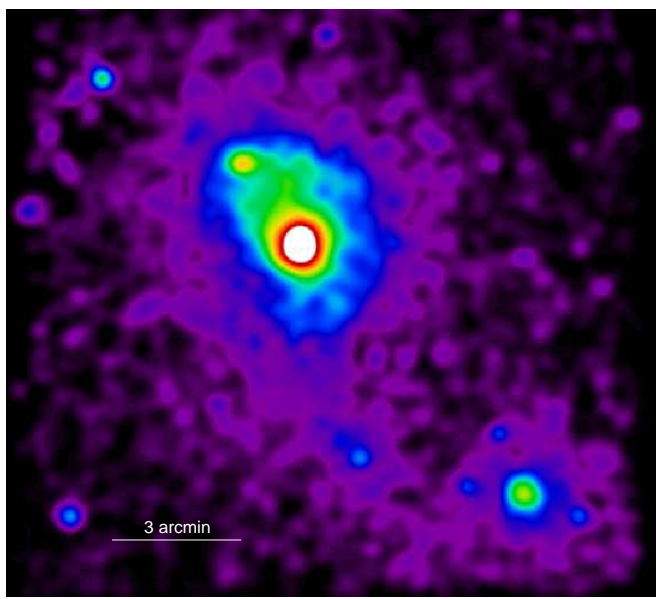}
  \caption{
    Chandra image of the cluster system A3016/A3017 in the 0.5--2.0 energy keV band. 
    The brightest component in the northern region is A3017 and A3016 is identified 
    with the component in the southwest. 
    The two cluster components are connected by a bridge of diffuse X-ray emission.
  }
  \label{fig:chandra}
\end{figure}

The cluster system was observed twice with Chandra (OBSID 15110, 17464) for 14.88 and 13.88 ks, respectively.
The data were taken with the ACIS-I detector in VFAINT mode. 
Standard data analysis was performed with CIAO 4.8.1 with calibration database (CALDB) 4.7.1. 
This includes flare cleaning by filtering light curves of source-free regions, VFAINT mode, gain, 
and charge transfer inefficiency corrections. 
The final exposure after cleaning is slightly less than 28~ks. 
We applied the same procedure to the ACIS Blank-sky data, which was used as our background estimate 
after adjusting the normalisation.
We used the 9.5--12 keV count rate to get the appropriate normalisation for the background exposure. 
We used a fluximage to obtain flux and exposure maps after background subtraction.
The resulting combined map is shown in Fig.~\ref{fig:chandra} in the 0.5--2 keV band.

For the spectral analysis we used the area well outside $r_{500}$ of the cluster as a background 
region in each dataset. 
We considered three background components: unresolved X-ray point sources, the local bubble, and the hot halo. 
The first component was modelled with a power law for a fixed index of 1.4, the second with a thermal model 
for a fixed temperature of 0.1~keV and the last one for a fixed temperature of 0.25~keV.
The amplitudes of these three background components were fitted in the background region of the data 
and were scaled for each source region accordingly.

Our optical survey of the cluster includes photometric images of the system taken with the Wide Field 
Imager (WFI) at the ESO/MPG 2.2m telescope at ESO La Silla in four optical bands (B,V,Rc,I). 
Total integration time was 60 minutes for each of the B, Rc, and I bands and 30 minutes for the V band. 
The V band dataset was discarded due to its low signal-to-noise ratio (S/N) and sparse sampling. 
The total field of view for all datasets is 0.3~deg$^2$.

\begin{figure}
  \includegraphics[width=\columnwidth]{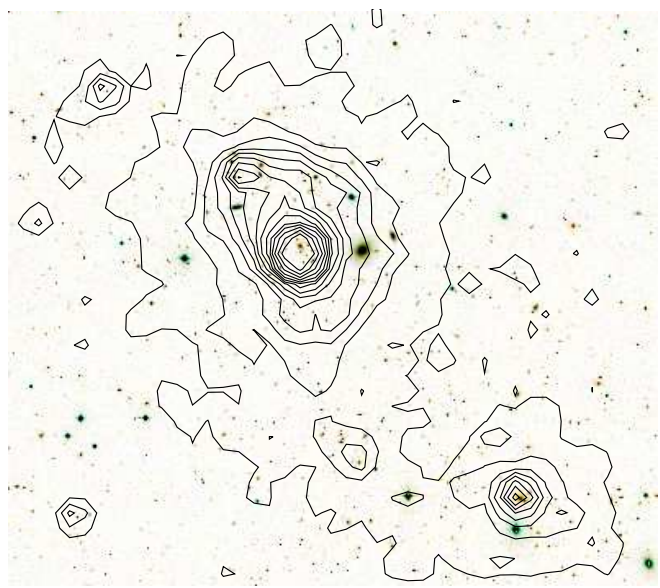}
  \caption{
    Surface-brightness contours of the Chandra superposed on an optical
    composite image based on the B, Rc, and I bands from the WFI data.
  }
  \label{fig:wfibig}
\end{figure}

The basic data reduction included bias subtraction, flat fielding, and masking of bad pixels and satellite tracks. 
Sky subtraction was performed by averaging the normalised images in which every source with a surface 
brightness of less than 28 mag/arcsec$^2$ was masked. 
This background model was then flux-rescaled to the individual exposures and was subtracted. 
General details for this method will be described in Kluge et al. (in prep.). 
The astrometric and photometric solutions were calculated using SCAMP~\citep{bertin06} and the final 
calibrated images were co-added using SWarp~\citep{bertin02}.
A composite image of the WFI data based on the three bands is shown in Fig.~\ref{fig:wfibig} with 
the surface brightness contours of the Chandra data.

We identified 8697 sources with SExtractor~\citep{bertin96} in the co-added Rc-band image.
These sources were sorted according to their stellarity indices and cluster member candidates 
were selected for those sources whose apparent brightnesses were between $17.7 < m_{Rc} < 22.0$ 
and $17.0 < m_{I} < 21.0$. 
We used two red sequences based on B-Rc and Rc-I to select cluster members applying the thresholds 
of $\pm$~0.15~mag and $\pm$~0.08~mag around the fitted red sequence, respectively. 
This results in a sample of 610 galaxies which belong to both red sequences.

\section{Large-scale morphology of the cluster complex}

Figure~\ref{fig:chandra} shows a Chandra X-ray image of the cluster complex in the 0.5--2.0 keV band. 
The image shows the two major sub-cluster components: the main cluster, A3017, in the upper left,
and A3016, which has a lower X-ray luminosity than A3017, in the lower right. 
The two clusters are connected by a bridge of X-ray emission. 

\begin{figure}
  \includegraphics[width=\columnwidth]{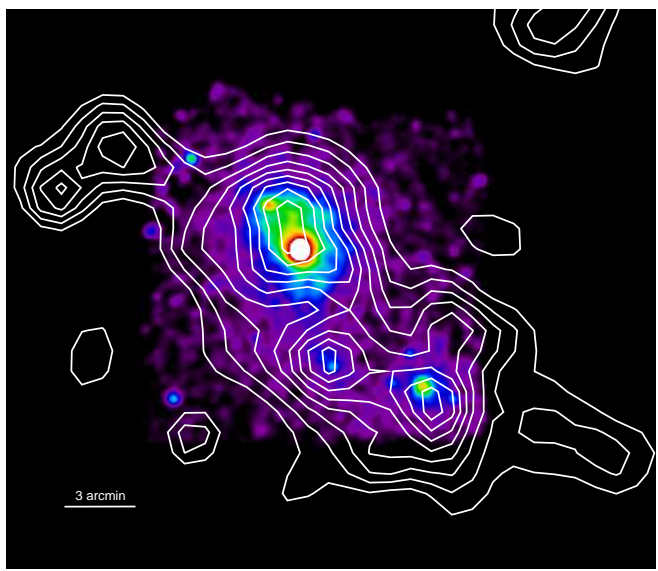}
  \caption{
    Distribution of the red-sequence galaxies in the RXCJ0225.9-4154 
    cluster region shown in white contours superposed on the Chandra image.
  }
  \label{fig:redseq}
\end{figure}

\begin{figure}[h]
  \includegraphics[width=0.45\columnwidth]{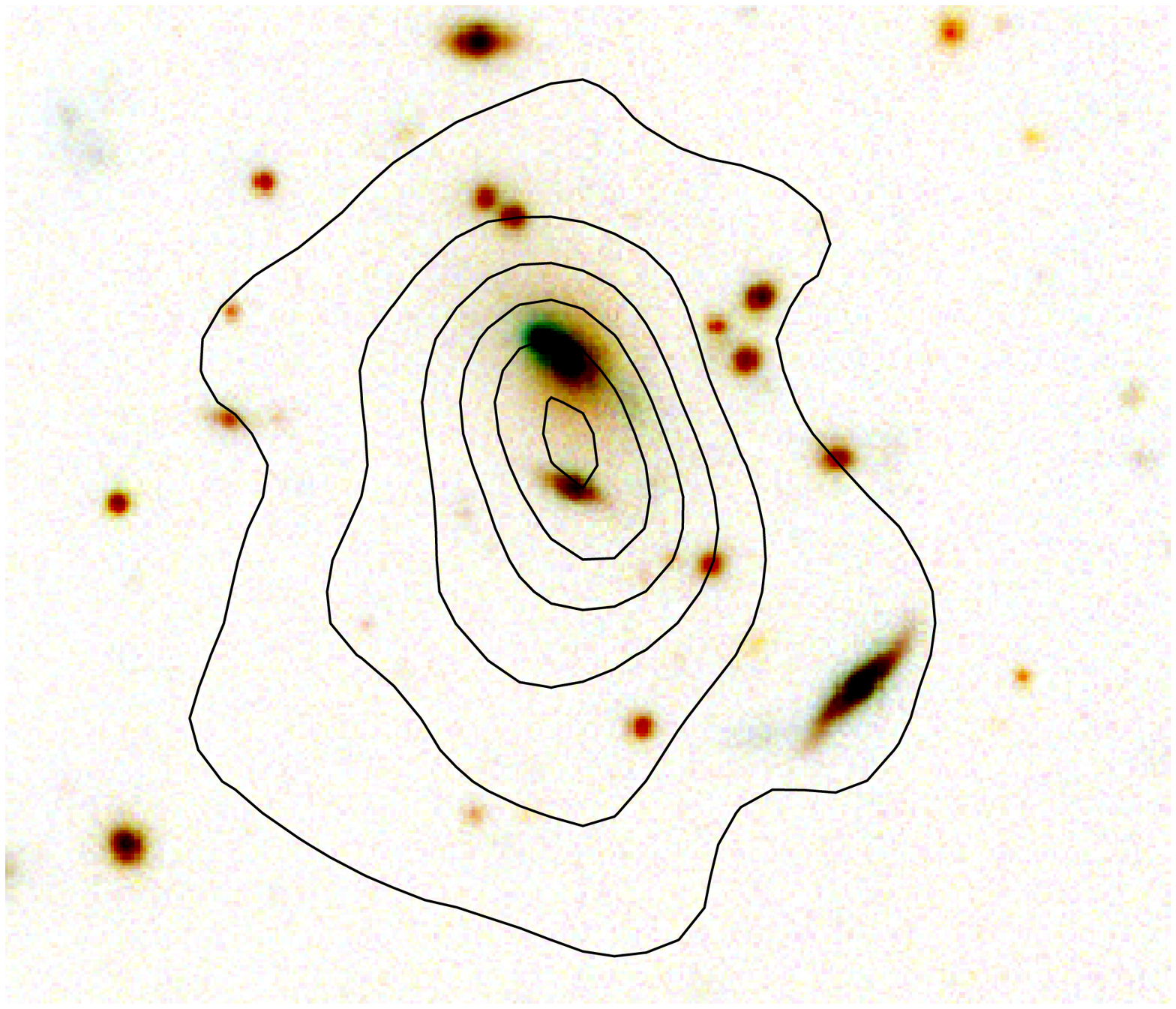}
  \includegraphics[width=0.45\columnwidth]{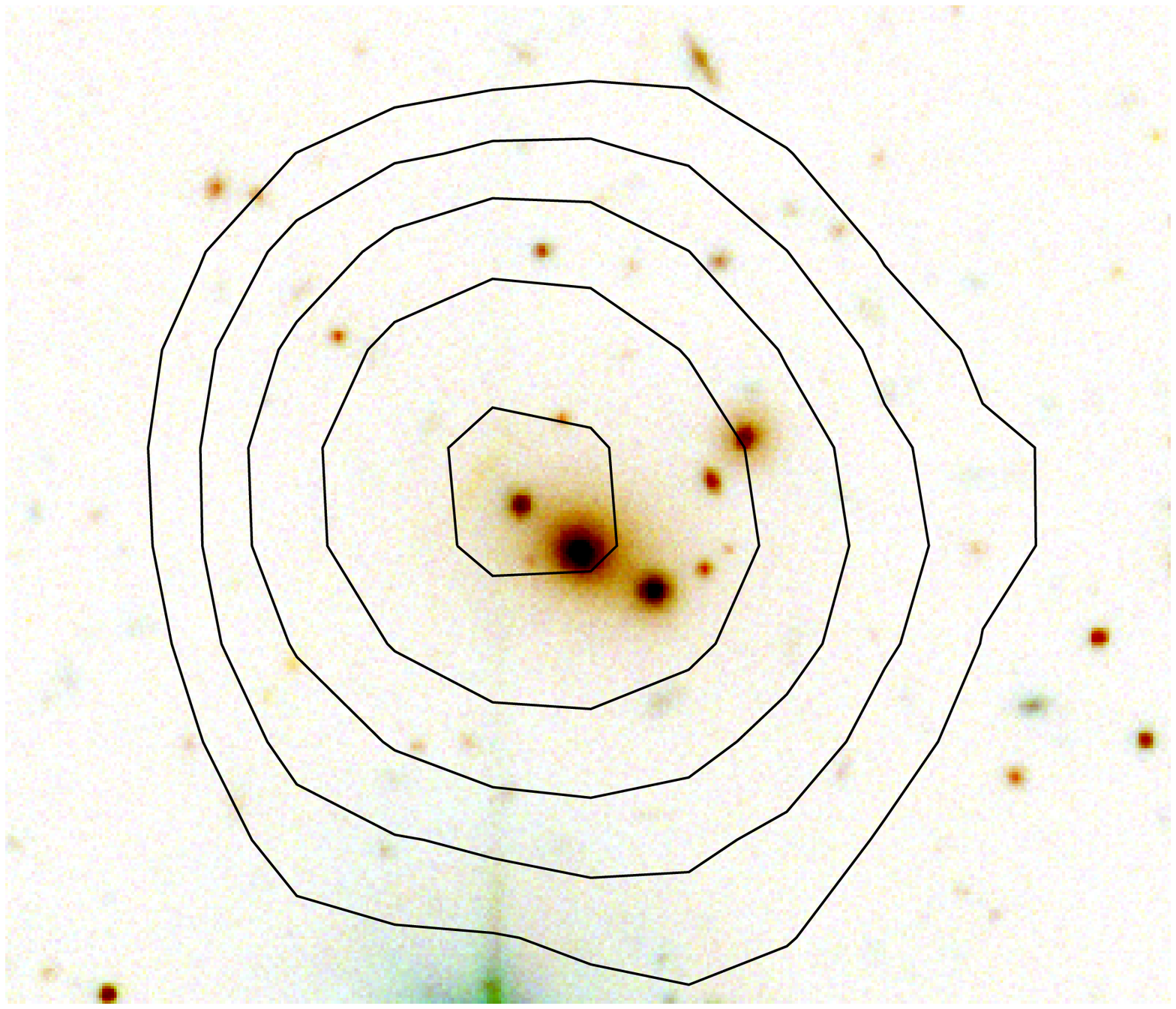}
  \includegraphics[width=0.45\columnwidth]{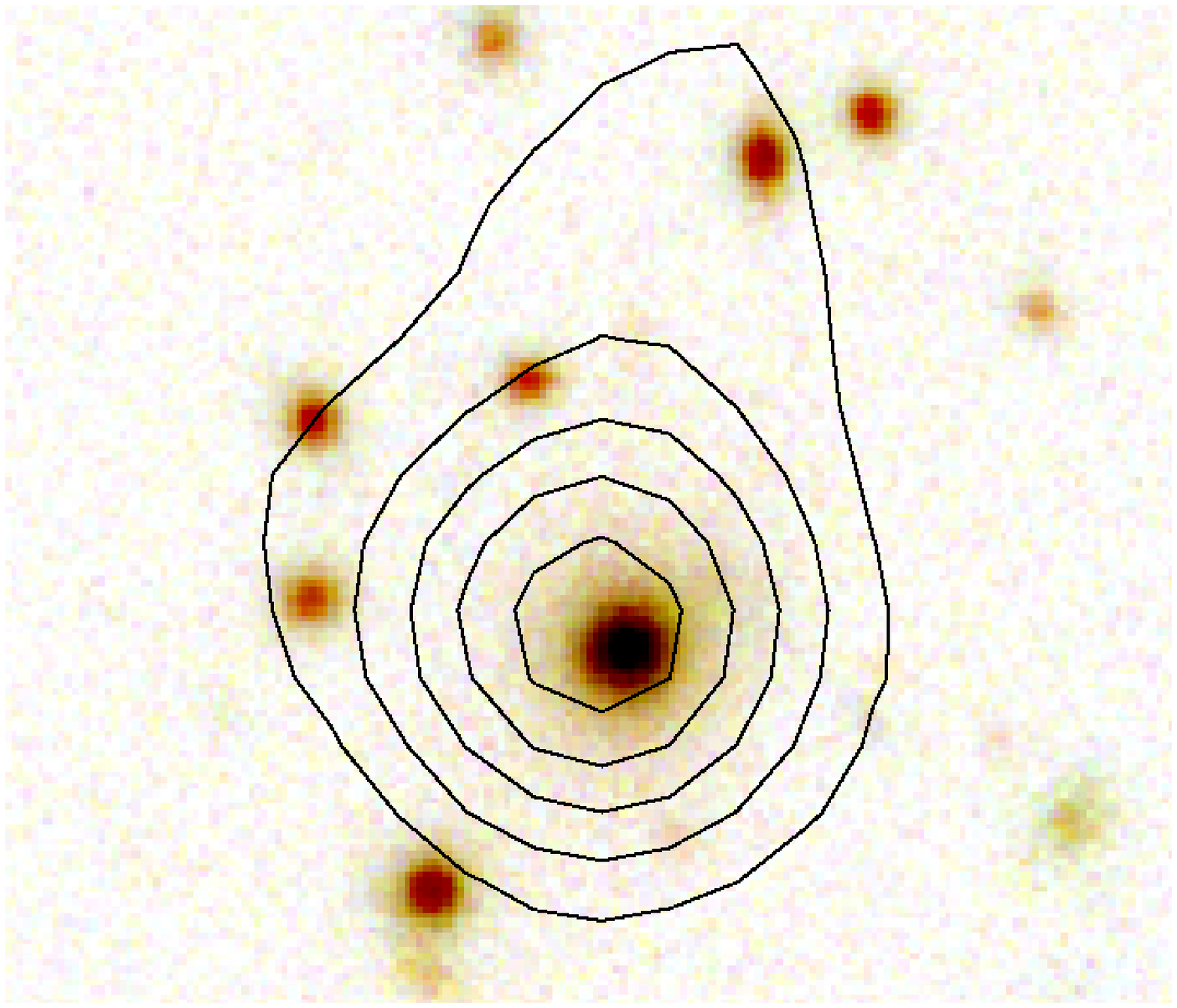}
  \hspace{0.7cm}
  \includegraphics[width=0.45\columnwidth]{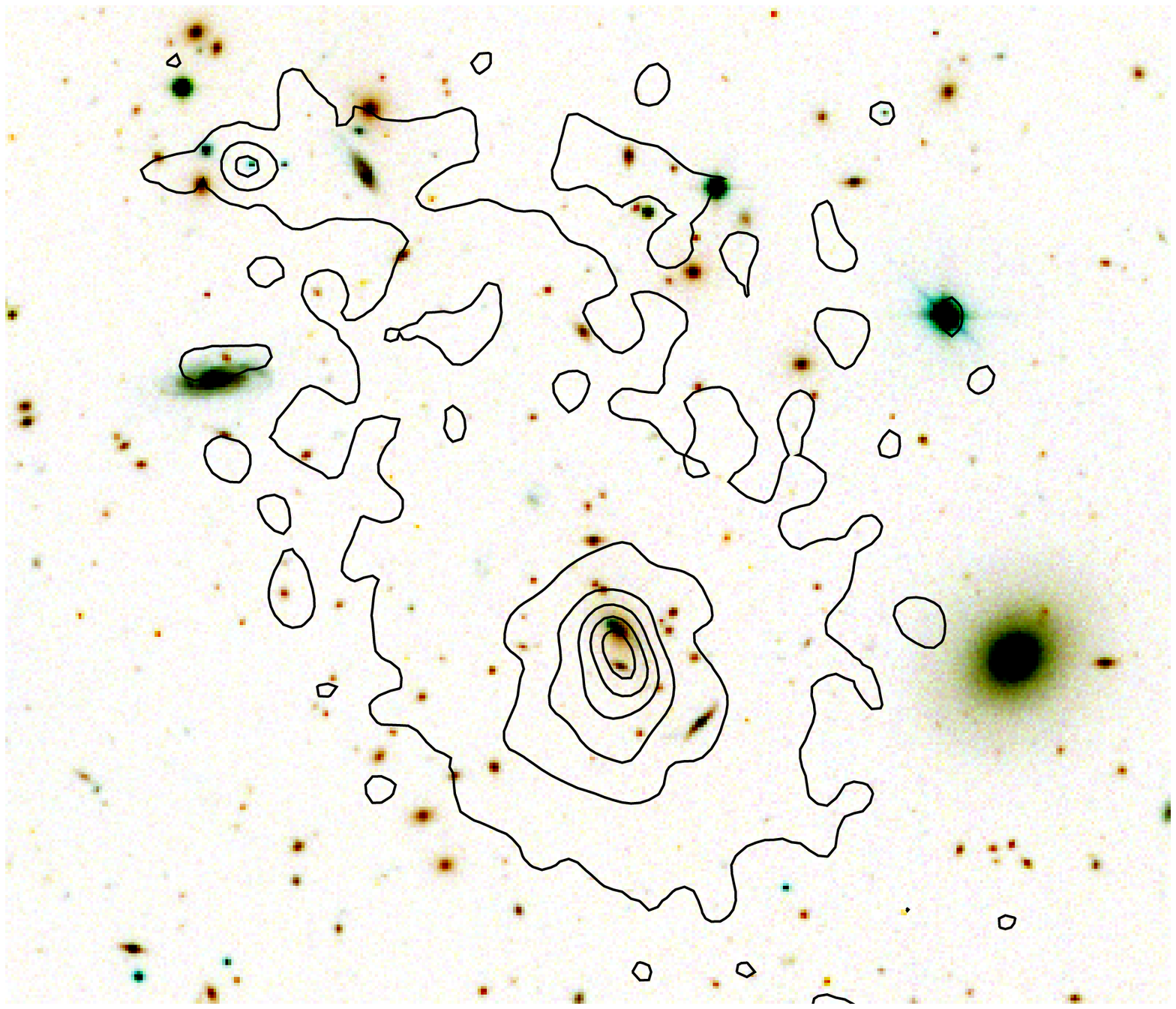}
   \caption{
     X-ray contours superposed on the optical image from the ESO/MPG Wide Field Imager.
     Shown are the central region of A3017 (top left), A3016 (top right), 
     the intermediate group (bottom left), and the northern clump (bottom right). 
     We note that the centre of the X-ray emission is also marked 
     in all first three figures by a bright early-type galaxy. 
   }
   \label{fig:wfi}
\end{figure}

A contour plot of the red-sequence galaxy distribution superposed on the X-ray image is shown 
in Fig.~\ref{fig:redseq}. 
The images show that the structure traced by the X-ray gas is embedded in  a large-scale structure filament. 
The galaxy distribution follows the orientation of the X-ray emission very closely and extends beyond it, 
displaying a galaxy filament which stretches in the northeast to southwest direction over at least 6~Mpc. 
In the southern part, the filament could have an even larger extent, since in the figure the filament ends 
at the edge of the WFI field-of-view. 
Both sub-clusters have a luminous, dominant elliptical galaxy near the X-ray peak. 
The galaxy redshifts show that we are almost looking from a perpendicular direction at the filament. 
The southern cluster has a velocity difference to the main cluster of about 650~km~s$^{-1}$ and the filament 
of 250~km~s$^{-1}$~\citep{foex17}.
Also, the distribution of the red-sequence galaxies confirms that the two Abell clusters are indeed associated.
This makes our interpretation of the system more straightforward than that given in~\citet{parekh}. 
In Fig.~\ref{fig:redseq} one can clearly see that the galaxy overdensities coincide with regions of 
high X-ray surface brightness.
This is also shown in the top two panels of Fig.~\ref{fig:wfi}, where we find giant galaxies near 
the X-ray maxima of the two clusters, typical brightest cluster galaxies (BCG).
We note that the largest galaxy in Fig.~\ref{fig:wfibig} and in the top-left panel of Fig.~\ref{fig:wfi}, 
at the distance of 1.5~arcmin from the centre of A3017 is a foreground galaxy, 6dF J0225450-415458 (z=0.0178).

Closer inspection of the X-ray surface brightness distribution shows, apart from the three main features, 
the bridge between the sub-clusters.
Within the bridge there is a compact source, centred on a luminous galaxy. 
The bottom-left panel of Fig.~\ref{fig:wfi} focusses on this feature, which clearly shows extended X-ray emission.
The surrounding ICM is also brighter and in the optical we note a clear galaxy concentration 
as shown in Fig.~\ref{fig:redseq}. 
Therefore this part of the bridge harbours a smaller group of galaxies with a central dominant galaxy, 
which seems to harbour its own coronal X-ray halo.

The X-ray halo of A3017, the main cluster of the system, is strongly elongated towards the northeast. 
At smaller radii from the cluster centre there is a bright feature that protrudes with its own X-ray peak.
The bottom right panel in Fig.~\ref{fig:wfi} shows that this clump of X-ray emitting gas has no luminous 
galaxy inside, thus it is not a freshly infalling X-ray luminous group.
A small point source embedded in this extended emission, which coincides in the optical with a compact 
bluish object, is most probably an AGN unassociated with the cluster.
Therefore the localised peak in the extended emission of the northern clump in A3017 is probably
due to this contaminating source.
At larger radii from the cluster centre to the northeast we see fainter X-ray emission that follows 
the extension of the galaxy filament in this direction. 
It is probably marking the connection of the cosmic filament to the cluster. 

Thus we note six features in the X-ray image, that we further
investigate in the following: 
(i) the main sub-cluster A3017 
(ii) the second sub-cluster, A3016
(iii) the X-ray luminous bridge between the clusters
(iv) the X-ray luminous galaxy group near the middle of the bridge
(v) the X-ray luminous clump to the northeast of A3017 
(vi) the extension of the X-ray surface brightness towards the northeast close to the direction
of the large-scale structure filament.
Features (i)-(v) are marked in Fig.~\ref{fig:tregs} and (vi) is 
indicated by the most northern box in Fig.~\ref{fig:bridge}.

\section{Cluster components}
\subsection{Properties of the cluster components}

\begin{figure}
  \includegraphics[width=\columnwidth]{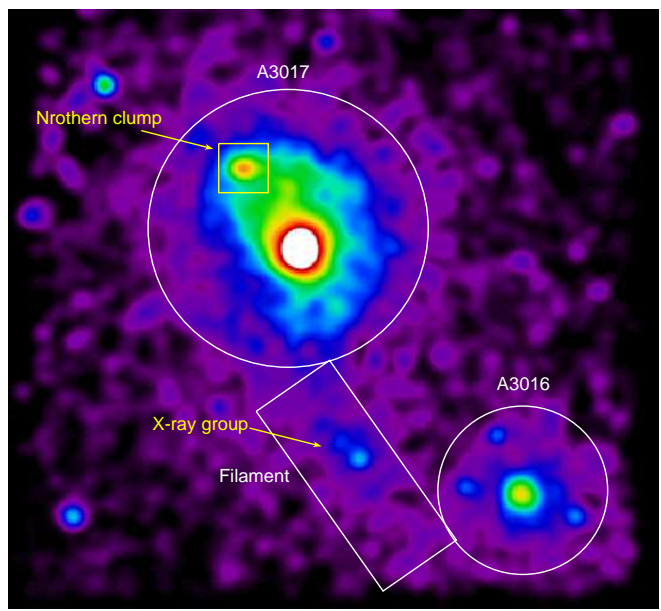}
  \caption{
    X-ray image of RXCJ0225.9-4154 with the three regions marked
    in white where the spectral temperatures were determined.
  }
  \label{fig:tregs}
\end{figure}

We used X-ray spectroscopic data to determine temperatures for different parts of the cluster complex 
to obtain further information about the different components.
Spectra were extracted in three regions of the cluster as defined by the white polygons in Fig.~\ref{fig:tregs}. 

%% v2
Fitting these regions, we find 7.05$^{+0.66}_{-0.59}$~keV for A3017 (T1), 
3.92$^{+0.86}_{-0.68}$~keV for A3016 (T2) 
and 4.14$^{+3.06}_{-1.38}$~keV for the area between the two clusters (T3).
The value of r$_{500}$ of A3017 is 5.8~arcmin, but the spectrum was extracted between 0.3 and 3 arcmin 
to avoid the southern interaction region and also the cool-core.
That of A3016 is 4.2~arcmin, and for the same reason as above, the spectrum was extracted inside the circular 
region of 2 arcmin.
The size of the T3 region is 2 by 5 arcmin$^2$.
The metallicities for A3017 and A3016 were also fitted yielding the values of  
0.19$\pm$0.1 $Z_{\odot}$ and 0.23$^{+0.38}_{-0.23}Z_{\odot}$, respectively.
The metallicity was fixed to 0.2 solar for T3 due to photon statistics.
The temperature fit in the T3 region has a very large uncertainty due to low photon statistics. 
Decreasing the metallicity to 0.1 solar results in the best-fit temperature of 3.36$^{+3.62}_{-1.64}$~keV.
The ICM in these regions may have a multi-temperature structure, but due to the relatively short exposure 
we cannot obtain more accurate information by either subdividing the regions into smaller bins or fitting 
more complex spectral models. 
Within the relatively large error budget the temperature distribution indicates that the filament 
between the two clusters has likely been heated due to their interaction.

\begin{figure}
  \includegraphics[width=\columnwidth]{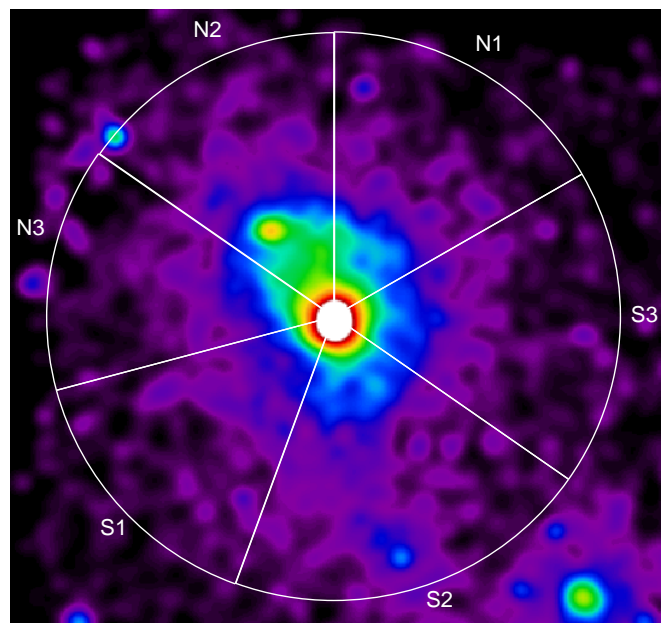}
  \caption{
    X-ray image of RXCJ0225.9-4154 with the six sectors in
    A3017 from which surface brightness profiles were extracted.
  }
  \label{fig:ssect}
\end{figure}

To estimate the masses of the two sub-clusters we also determined X-ray surface brightness profiles 
of these components. 
Since A3017 shows major disturbances of an azimuthally symmetric structure, we sub-divided the 
cluster data into six sectors as shown in Fig.~\ref{fig:ssect}. 
The northern sectors N1 and N2 are clearly disturbed by the accretion of material from the northern filament.
The sector S2 shows an extension into the bridge. 
Only the sectors S3 and N3 seem to reflect the shape of the cluster undisturbed by recent accretion. 

A close inspection of the sector S1 in the X-ray image of Fig.~\ref{fig:ssect} shows that the surface 
brightness contours are squeezed at radii between about one and two~arcmin. 
Since we suspect that this could have been caused by an interaction of the two sub-clusters, 
we subject this sector to further investigation. 

\begin{figure}
  \includegraphics[width=\columnwidth]{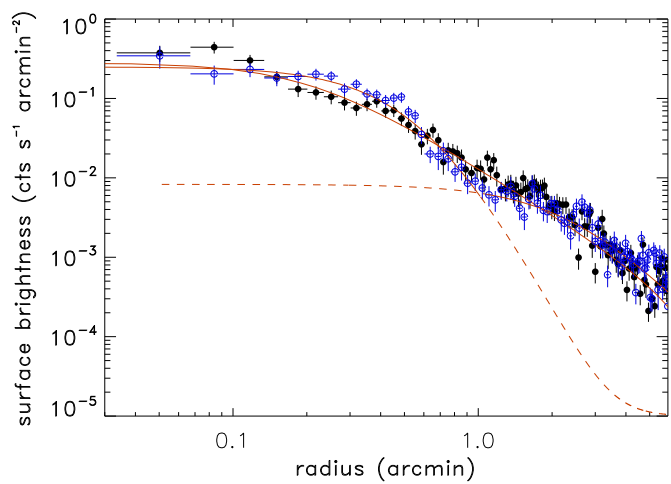}
  \caption{
    X-ray surface brightness profiles of the sectors S1 (blue)
    and S3 (black) of the A3017 sub-cluster. 
    The profiles were fitted with $\beta$-models as shown by the solid lines.
    For the sector S1 the inner and outer profiles were fitted by 
    two separate $\beta$-models.
  }
  \label{fig:sbprof}
\end{figure}

In Fig.~\ref{fig:sbprof} we show the average surface-brightness profile in sectors S3 and S1. 
The surface-brightness profile in sector N3 is very similar to that of S3. 
We clearly note the deficit of X-ray-emitting gas in sector S1 compared to sector S2 in the region 
from about 0.7 to 2 arcmin.  

From the X-ray surface-brightness distribution and the measured ICM 
temperatures we estimated the total and the gas masses of the different 
components of the merging cluster complex. 
The mass estimates were determined assuming hydrostatic equilibrium of 
the ICM and approximate spherical symmetry of the cluster components.
Due to the sparsity of the data we keep the spatial modeling relatively 
simple, as most of the uncertainty in the mass originates from the 
uncertainties on the ICM temperature determination. 
Therefore, we modelled the surface-brightness distribution with a $\beta$-model
~\citep{cavaliere76} and made sure that the best fit is a fair 
description of the data. 
For the temperature profile we used two solutions, which bracket the typical
temperature distributions in clusters outside the core region, as found for 
example in our REXCESS survey of a representative sample of X-ray clusters
from X-ray sky surveys~\citep{boehringer07}. 
We use an isothermal profile and profiles with polytropic indices of 1.2
and 1.1, which are characterised by $T(r) \propto \rho(r)^{0.2}$. 
The polytropic temperature profiles are normalised such that the emission-measure-weighted average temperature outside the core is equal to the 
measured bulk temperature. 
In the case of A3017 we excised the core of radius 0.3 arcmin to determine
the bulk temperature as indicated at the beginning of this section. 
For the overall result of the mass profile we evaluated the best fit for
a polytropic index of 1.1 and the minimum and maximum of the combinations
of all extreme density and temperature profiles.

\begin{figure}
  \includegraphics[width=\columnwidth]{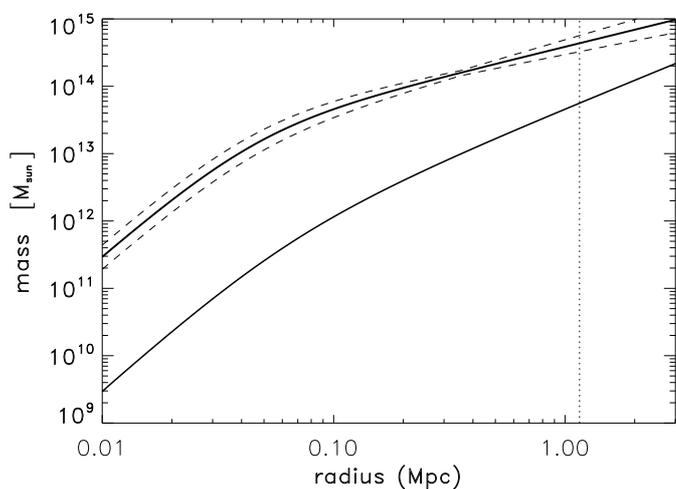}
  \caption{
    Total (upper solid line) and gas mass profiles 
    (lower solid line) for A3017, the main component of
    the cluster complex. 
    The profile was determined from the relatively undisturbed sector, S3, 
    in the western side of this cluster.
    The two dashed lines indicate the uncertainty of the total 
    mass determination.
  }
  \label{fig:massprof}
\end{figure}

For A3017 we obtained the following results from the best-fitting $\beta$-model 
to the surface-brightness profile of sector S3 (with similar values for N3). 
A resulting core radius of 0.18 arcmin and a $\beta$-parameter of 0.48
implies a cluster mass of $M_{500} = 3.9^{+1.9}_{-1.2} \times 10^{14}$~M$_{\odot}$ 
with the above quoted values for the bulk temperature of the ICM of this 
component. 
For the gas mass of A3017 we obtained a value of $M_{500} = 5.6 \times 10^{13}$~M$_{\odot}$.
Figure~\ref{fig:massprof} shows the resulting mass and gas mass profiles 
for A3017.

From the difference of the two surface-brightness profiles for the sectors 
S1 and S3 we find that approximately $3 - 4 \times 10^{11}$~M$_{\odot}$ of gas is 
missing in sector S1 in the radial range from about 200 to 400~kpc, 
compared to what we consider as the undisturbed cluster. 
This amounts to about one third of the gas mass inside 400~kpc in sector S3.
At larger radii we do not note such a deficit. 

The southern component, A3016, is less disturbed than the main cluster. 
Here it is the northeast part of the cluster that overlaps with the bridge 
between the clusters. 
Therefore we used the southwest half of the cluster to obtain an undisturbed
reference surface brightness profile, yielding a core radius of 0.13 arcmin 
and a $\beta$-parameter of 0.46. 
The resulting total mass based on the temperature determination quoted above
is $M_{500}=1.3^{0.9}_{-0.5} \times 10^{14}$~M$_{\odot}$ with a gas mass of 
$M_{500} = 1.6 \times 10^{13}$~M$_{\odot}$.

As a first consistency check we may look at the gas mass fractions 
of the two sub-clusters, which results in $10 - 16\%$ for A3017 and 
$9 - 14\%$ for A3016. 
This covers the expected values for the gas mass fraction for high mass 
clusters found to be in the range of about $8 - 15\%$  in the 
studies by~\citet{vikhlinin06} and~\citet{pratt09}, for example.
We can also compare to the mass estimate from our X-ray luminosity 
to mass relation from the REXCESS study~\citep{boehringer07,pratt09}
which yields an X-ray luminosity for A3017 
of $L_{500} \sim 4-5 \times 10^{44}$~erg~s$^{-1}$ and a mass of $6.2 \times 10^{14}$~M$_{\odot}$, and for 
A3016 of $L_{500} \sim 0.7 \times 10^{44}$~erg~s$^{-1}$ 
with an estimated mass of $1.8 \times 10^{14}$~M$_{\odot}$.

%% v2
Similarly we can use the X-ray temperature--mass relation to estimate the sub-cluster masses 
(e.g. Arnaud et al. 2005).
For the above quoted temperatures we obtain the following mass estimates: 
$M$ = 7.0$\pm$1.3$\times$10$^{14}$ M$_\odot$ for A3017 and 
$M$ = 2.6$^{+1.5}_{-0.8}\times$10$^{14}$ M$_\odot$ for A3016. 
While consistent within the 1$\sigma$ uncertainties, the masses estimated this way are higher 
than the values derived above.
There may be two reasons for this; the core-excised temperatures are higher than those for the whole 
cluster regions, and the clusters used in Arnaud et al. (2005) to construct the scaling relations
are preferentially regular, and therefore are expected to have a slightly different scaling relation 
than this disturbed system (see e.g.~\citet{chon12_subs,chon17}).
The coronal halo of the central galaxy of the group in the middle 
of the bridge has an X-ray luminosity of 
$L_X = \sim 2.3 \times 10^{42}$~erg~s$^{-1}$ in the 0.5--2 keV 
band, typical for the hot gas in massive ellipticals in groups. 

The northern clump has an X-ray luminosity of about 
$2.2 \times 10^{43}$~erg~s$^{-1}$ in a region of 
about 250 by 250~kpc$^2$ after removing the contribution 
from a point source.
This is consistent with a smaller infalling group of galaxies, but the centre of this region is not 
marked by a giant elliptical as we would expect for a group. 
Therefore, either the gas has been separated from the associated galaxy or the gas is not from a group but 
is a sub-structure feature of A3017. 

\subsection{Internal structure of A3017}

\begin{figure}
  \includegraphics[width=\columnwidth]{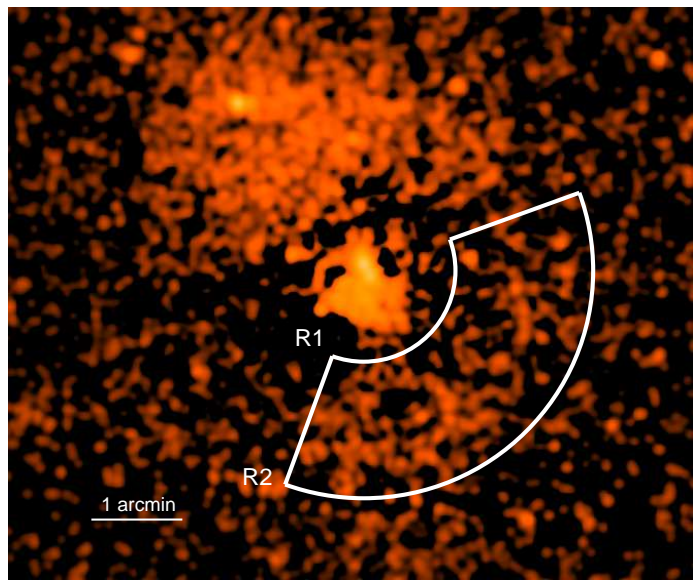}
  \caption{
    X-ray residual map of A3017 after subtracting the best-fit
    $\beta$-model based on the sector S3 in Fig. 6. 
    The region bounded by two radii, R1 and R2, corresponds
    to the spiral arm identified by~\citet{parekh}.
    In our analysis R1 and R2 correspond to the radii identified 
    as the cold front and the shock front in Fig.~\ref{fig:s2prof}.
  }
  \label{fig:resid}
\end{figure}

\begin{figure}
  \includegraphics[width=\columnwidth]{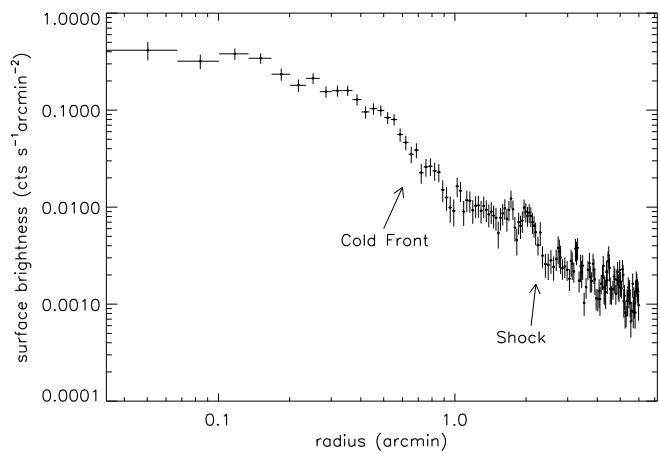}
  \caption{
    Surface brightness profile of the sector S2 as defined in Fig.~\ref{fig:ssect}.
  }
  \label{fig:s2prof}
\end{figure}

A detailed analysis of the X-ray surface-brightness distribution and the ICM temperature
inside A3017 reveals further interesting features.
\citet{parekh} described a spiral-arm structure in the surface brightness in the western 
outskirts of A3017 as a cold front with its gas temperature of 6~keV. 
Using our best fit $\beta$-model for A3017 described in Section 4, we obtained the X-ray residual 
map of A3017 in Fig.~\ref{fig:resid}.
We confirm the excess of the X-ray gas at a similar location in Fig. ~\ref{fig:resid} to that in 
Fig. 1(b) or Fig. 3 of~\citet{parekh}.
A close inspection of our image shows that, rather than a spiral, this is an arc-like structure with an almost 
constant radius to the centre.
Our estimation of the gas temperature is 8.9$^{+3.3}_{-1.7}$~keV.
To determine if this structure corresponds to a sloshing front, we extracted temperatures on both 
sides of the arc region at radii R1 and R2 in Fig.~\ref{fig:resid}.
These radii were guided by two breaks in the surface-brightness profile of the sector S2 shown 
in Fig.~\ref{fig:s2prof}.
We find that the temperature of the region below the radius R1 is 4.1$^{+0.5}_{-0.4}$~keV and that 
beyond R2 is 5.8$^{+4.3}_{-1.9}$~keV.
The temperature jump at R2 does not support the suggestion by~\citet{parekh} that the western
structure is the result of tidal sloshing. 
The observations are more consistent with a shock front at R2.
We rather have an indication of a cold front close to the core at R1, a radial region where most cold 
fronts have been observed (e.g.~\citet{ghizzardi10}).
It is interesting that we find the signature of a shock front in the cluster region that is essentially opposite
to the massive accretion of material from the filament in the northeast.
We may therefore see the forward shock in response to the compression of the infalling material.

\section{Properties of the filaments}
 
\begin{figure}
  \includegraphics[width=\columnwidth]{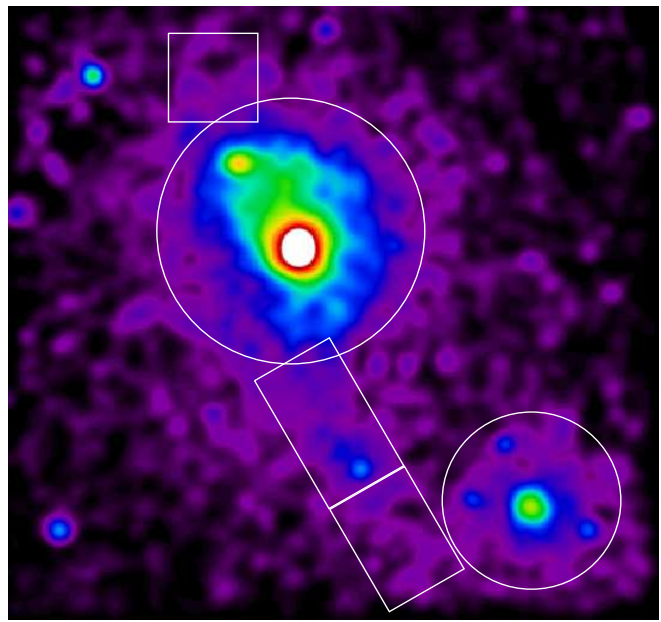}
  \caption{
    X-ray image of RXCJ0225.9-4154 with the two study regions
    of the X-ray luminous bridge marked by the two connected 
    white boxes and the connection to the filament marked
    by one white box. For the two sub-clusters a three arcmin radius 
    is shown for A3017 and a two arcmin radius for A3016.
  }
  \label{fig:bridge}
\end{figure}

To analyse the physical conditions in the bridge area, we selected two study regions, as marked in 
Fig.~\ref{fig:bridge}, with sizes of 4 by 1.9 arcmin for the northern region and 2.5 by 1.9 arcmin in the south.
In both regions taken together, we find a significance of excess X-ray emission of about 18$\sigma$ after 
subtracting the two cluster emission models from the data.
Inside the two regions we have modest surface-brightness variations and the northern region is about a factor 
of 1.8 brighter than the southern one. 
We also note that the northern region is brighter than the cluster emission of A3017 outside about 3~arcmin 
(0.6~Mpc) and the southern region is brighter than A3016 outside about 2~arcmin (0.4~Mpc). 

For the northern region we find an average surface brightness 
of about $S_X \sim 2 \times 10^{-3}$~cts~s$^{-1}$~arcmin$^{-2}$
and a total luminosity of about $1.9 \times 10^{43}$~erg~s$^{-1}$
in the 0.5 -- 2 keV band. 
The region has an extent of about $0.85 \times 0.4$~Mpc$^2$. 
If we assume that it extends for about 0.5~Mpc in depth, we can 
determine the average electron density in this part of the bridge to be 
$n_e \sim 8.2 \times 10^{-4}$~cm$^{-3}$ and the total gas mass 
$M_{gas} \sim 3.4 \times 10^{12}$~M$_{\odot}$. 
For the southern region, with an area of $0.53 \times 0.4$~Mpc$^2$ 
and a total luminosity of $0.74 \times 10^{43}$~erg~s$^{-1}$
we find an electron density of $n_e \sim 6.4 \times 10^{-4}$~cm$^{-3}$  
and a total gas mass of $M_{gas} \sim 1.7 \times 10^{12}$~M$_{\odot}$.
Thus the total mass of gas contained in the selected region of the
bridge amounts approximately to $5 \times 10^{12}$~M$_{\odot}$
and there may well be at least a similar amount of gas outside this
bright central area, meaning that the total gas mass in the bridge 
could well be $10^{13}$~M$_{\odot}$. 
An inhomogeneous gas distribution in the analysed region will lead 
to an overestimate of the gas mass.
This overestimate is certainly much less than a factor of two,
since this factor would require
the density fluctuations to have unexpectedly large amplitudes.
A source of uncertainty in determining the gas mass is the unknown
extent of the ICM in the bridge along the line-of-sight. 
A factor of two change in the line-of-sight length corresponds to 
a change in the the gas mass of only the square root of two.

For the northern extension of the X-ray surface brightness, as marked by a box
with a size of 2.5 by 2.5 arcmin$^2$ in Fig.~\ref{fig:bridge}, we find
a significance of excess X-ray emission of 8.7$\sigma$ after subtracting the
cluster emission modelled by an azimuthally symmetric $\beta$-model from the data.
The surface brightness in this region is about $1.36 \times 10^{-3}$~cts~s$^{-1}$~arcmin$^{-2}$.

The surface brightness has a clear radial gradient. 
If we take average properties we find a gas density of 
$n_e \sim 6.8 \times 10^{-4}$~cm$^{-3}$
and a total mass of about $M_{gas} \sim 2.3 \times 10^{12}$~M$_{\odot}$,
which is 14\% of the gas mass of A3017.

\section{The cluster merger and the interaction with the filaments}

The observed morphology of the cluster system suggests two different 
interpretations of the evolutionary state of the cluster merger. 
The system could be post-merger, where the sub-cluster A3016 has 
passed A3017 in a close fly-by and gas was ripped off from both systems, 
which now forms the bridge between them. 
Alternatively, the system is at the stage of the first collapse and 
the material of the bridge is another overdensity within the large-scale 
filament hosting a galaxy group near its centre.  

The steep surface-brightness profile and the gas mass missing in sector S1 
of A3017 is very suggestive of a scenario in which A3016 has passed A3017 
on the eastern side and has ripped out gas where it is now missing. 
The gas bridge is then trailing A3016, forming an arc following the
trajectory of A3016.
A3016 may have nearly reached a turning point from which it falls back 
on A3017. 
There is a major caveat with this interpretation.
As we have seen above, the gas mass that is missing in sector S1 with 
$3 - 4 \times 10^{11}$~M$_{\odot}$ falls far short of the gas mass of 
the bridge, which is approximately $0.5 - 1 \times 10^{13}$~M$_{\odot}$. 
It is also plausible that more gas could have been stripped off 
from the outer parts of both clusters. 
An approximately 13\% loss of the gas of\  A3017 would  be sufficient,
but most of this gas would be in the cluster outskirts and would 
originally have a low gas density. 
The gas in the bridge has a higher density and lower entropy than the 
gas from larger cluster radii, and it is difficult to imagine a process 
by which the gas is ripped off from the clusters and then compressed 
to lower entropy by the merger event.
In addition we may conclude that a possible first merger could not have 
been very dramatic, since both clusters have still retained most of 
their gas inside at least about 0.7$r_{500}$.
In the case of a post-merger scenario this probably implies a 
sufficiently large impact parameter.

In the second scenario, the material of the bridge comes from a separate overdense region between the clusters. 
In this case we would expect the gas in the bridge to be colder than the gas in the thermalised clusters. 
However, the temperature measurement from the X-ray spectrum shows a relatively high temperature for 
the bridge gas. 
This implies that the gas in the bridge must also have been shock-heated. 
The bridge region is just about covered by the two $r_{500}$ radii of A3107 and A3016. 
The accretion shocks of both clusters are expected to be located well outside these radii. 
Therefore, it is very plausible that the bridge material has been affected by the outgoing shocks 
from the clusters.
The existence of the group within the bridge may be a further indication of a separate origin.
It would otherwise be difficult to explain the formation of this group and its BCG in the stripping tail of A3017.
The group itself may have contributed a significant fraction of the X-ray-emitting plasma in the bridge,
and a previous interaction of the group with A3017 could be another reason for the high temperature in
the filament.

With this evidence, that is, mostly the fact that it is difficult to account for
the large gas mass in the bridge within a post-merger model, our second 
scenario seems to be more consistent. 
A final decision on which of the two pictures is most likely should await better 
X-ray observational data which would allow further diagnostics.
If the filamentary gas was heated by outgoing shocks, we would expect 
a decreasing temperature profile with radius. 
Also, a comparison of the metallicity of the gas in the clusters with that 
of the bridge would help to resolve the origin of the bridge gas. 
A metallicity in the bridge very different from that of the clusters 
would point to a separate origin.

The structures seen in X-rays to the north of A3017 indicate further significant mass accretion.
From the X-ray luminosity of the northern clump of A3017, we can infer that a total mass 
of at least 8$\times 10^{13}$~M$_{\odot}$ is associated in this accretion event.
The X-ray extension into the northern filament at large radii involves a gas mass of 
at least 2~$\times~$10$^{12}$~M$_{\odot}$  which implies a total mass accretion at this cluster 
distance of at least ten times this mass.
Therefore we are witnessing an extremely active region of gravitational collapse at a knot 
of the cosmic web.
The six distinct structures of this cluster complex are expected to form one very massive cluster 
in about three to four gigayears.

\section{Summary}

We used  Chandra observations to study the galaxy cluster system RXCJ0225.9-4154 in which two Abell clusters are connected by a bridge of X-ray emitting gas. 
This object is rich with interesting X-ray features, six of which we analysed in detail.
Also supported by optical photometric data, this system is embedded in a large cosmic filamentary 
structure through which we are witnessing a process of mass accretion.
We give a brief summary of four important points: A3017 and A3016 form one system being at the same 
redshift together with the X-ray bridge between them. 
The ICM structure of A3017 shows internal disturbances with an indication of a shock front in the west 
at a radius of about 430~kpc and a cold front at the edge of the cool core.
The X-ray-emitting filament between these two clusters is likely to be shock-heated and harbours 
an X-ray galaxy group with a clear BCG. 
Our X-ray studies suggest two different interpretations of the evolutionary stage of the cluster merger.
Based on the amount of gas present in the bridge region it is more likely that we are witnessing an event 
prior to the major merger of A3017 and A3016.
We await deeper X-ray data to fully understand the details of the formation process of this system. 

\begin{acknowledgements}
We thank the referee for constructive comments, which helped to improve the paper.
HB and GC acknowledge support from the DFG Transregio Program TR33 and the Munich Excellence Cluster 
"Structure and Evolution of the Universe".
This research made use of the NASA/IPAC Extragalactic Database (NED), which is operated by the Jet 
Propulsion Laboratory under contract by NASA.
\end{acknowledgements}

\footnotesize{
  \bibliographystyle{aa}
  \bibliography{0225} % bib creates bbl file.
}

\end{document}